# REGULATION SIMULATION

**Philip Maymin**[1]
*NYU-Polytechnic Institute*
*Email: phil@maymin.com*

**ABSTRACT**

A deterministic trading strategy by a representative investor on a single market asset, which generates complex and realistic returns with its first four moments similar to the empirical values of European stock indices, is used to simulate the effects of financial regulation that either pricks bubbles, props up crashes, or both. The results suggest that regulation makes the market process appear more Gaussian and less complex, with the difference more pronounced for more frequent intervention, though particular periods can be worse than the non-regulated version, and that pricking bubbles and propping up crashes are not symmetrical.

---

[1] Address: Professor Philip Maymin, Department of Finance and Risk Engineering, NYU-Polytechnic Institute, Six MetroTech Center, Brooklyn, NY 11201, USA. Phone: 718.260.3175. Fax: 718.260.3355. Email: phil@maymin.com.

I am grateful to Stephen Wolfram and Jason Cawley for many helpful comments during the 2007 *A New Kind of Science* Summer School where I first developed the minimal model of financial complexity used in this paper. I am also grateful to the anonymous referees of this journal for their valuable comments.

## I. INTRODUCTION

Should the government regulate the market?

The appeal of regulatory intervention is that the government might be able to costlessly mitigate drastic declines, thus increasing returns, lowering volatility, and potentially reducing both the negative skewness and the kurtosis (fat tails). But how would we know without first testing the regulations on an otherwise free market?

The fear of regulatory intervention is that the government may create an illusory market that appears to be a safe investment while actually hiding a ticking time bomb that can explode with far greater devastation than would have occurred had the government never intervened at all. But how would we know whether a particular crash would have happened without the regulation without having an identical but unregulated market to compare it with?

I introduce a framework for examining such issues by simulating a deterministic price process with a regulatory overlay. This approach allows us to ask infinite what-if questions and compare results across different choices.

The deterministic price process that forms the basis of the simulation is the minimal model of financial complexity that I have described in earlier work. It assumes a representative investor trading a single market asset based solely on its past market movements. Regulations are incorporated as overrides on the price process that would have resulted had the investor been allowed to trade freely.

Here, I show that the first four moments of simulated returns from this model are similar to the first four moments of returns observed on European markets. Then I explore the consequences of a variety of regulatory regimes to determine the impact on the market process.

## II. LITERATURE REVIEW

The hypothetical effect of financial regulation is difficult to assess using historical data because we do not know how the market would have evolved had regulation been different. There are several approaches to this problem.

One could simply argue from psychological principles how banks and financial companies ought to be regulated, c.f. Avgouleas (2009), but such an approach does not answer the question of what exactly would have happened otherwise; it merely makes the case that certain laws are better than others.

One could use econometric analysis on past regulatory changes to attempt to isolate the effects of unanticipated regulatory changes, c.f. Quinn (1997), but with this approach one is necessarily limited to the actual historical regulatory record.

One could assume a particular stochastic process for the evolution of market prices, but this would be a circuitous route as the impact of regulation would have to be indirectly posited as an effect on the parameters rather than a direct effect on the market price process. Furthermore, standard stochastic models often have independent increments, meaning that without further domain-specific assumptions, regulation by itself is inherently unable to dramatically or chaotically alter the process at the point of intervention, an unnecessary restriction.

One could run experimental markets with imposed regulatory rules and compare the effects to experimental markets without the regulations, c.f. Gerding (2007), but the drawback is the expense and time of continually running various regulatory conditions. Furthermore, such an approach is best used to test a hypothesis one has already established, rather than to try out different possibilities.

The approach I take here is to use simulation of a particular deterministic market process. This approach allows an exploration of possibilities and provides a framework for evaluating the effects of regulation.

With this framework, we can answer such questions as: does piercing bubbles have an exactly symmetrically opposite effect to propping up crashes? What happens if the government does both? How do realized market dynamics compare across different regulatory regimes?

I use the minimal model of financial complexity, described in Maymin (2008). In this deterministic model, a single representative investor trades a single market asset. I focus only on the simplest case of one investor, one asset, two possible actions (buy and sell), and two possible states, which can be loosely thought of as emotional states of the investor such as "optimistic" or "fearful." The representative investor's daily decision to buy or sell is modeled as a two-state iterated finite automaton over the preceding *w* days, where *w* is the lookback window. Out of all $(2 \cdot 2)^{2 \cdot 2} = 256$ possible rules, only one, canonically numbered rule 54 following Wolfram (2003), generates complexity in the resulting price series starting from an initial condition of *w* consecutive UP moves. Further, rule 54 generates complexity for almost any lookback window *w*. Because it is a unique, simple rule that generates complexity, it can be referred to as the minimal model of financial complexity.

## III. DATA AND METHODOLOGY

In the remainder of this paper, we will explore this minimal model with a lookback window of $w = 22$ ticks. Such a lookback window has the advantage of generating a long cycle; it takes 4,194,303 ticks before the price series inevitably cycles. Since the input to the model is only the preceding 22 ticks, and there are only $2^{22} = 4{,}194{,}304$ possible 22-length sequences of UP and DOWN, the minimal model has essentially the maximal cycle length: almost every 22-length sequence of UPs and DOWNs is traversed between any two repeats of the same initial conditions.

Because we will be looking at the effects of regulations on runs, we will use an initial condition of alternating UPs and DOWNs, starting with an UP move. This merely has the effect of shifting the price series to start from a slightly different point.

If the market process is a sequence of UPs and DOWNs, assume that the regulatory action of government consists of reversing a market movement whenever it so decides. The two primary examples we'll consider are regulations aimed at pricking bubbles and propping up crashes. In particular, the government can prick a bubble by reversing an UP market trend following *n* consecutive UP movements, or it can prop up a crash by reversing a DOWN market trend following *n* consecutive down movements, or it can do both. (When it does neither, the results are exactly the same as in Figure 2, described below).

## IV. EMPIRICAL RESULTS

Figures 1 and 2 show the rolling mean, volatility, skewness, and kurtosis of 256-day returns for, respectively, the major European markets and the minimal model with a 22-tick lookback window with days composed of 2,048 consecutive ticks.

**Figure 1**. **Moments of Major European Indices.** The four graphs below show the rolling 252-day mean, volatility, skewness, and kurtosis for the EuroSTOXX 50, CAC, DAX, FTSE 100, and SMI. The summary statistics in curly braces are the minimum, 10% quantile, median, 90% quantile, and maximum across all data points for all indices. Note that the moments differ considerably across time.

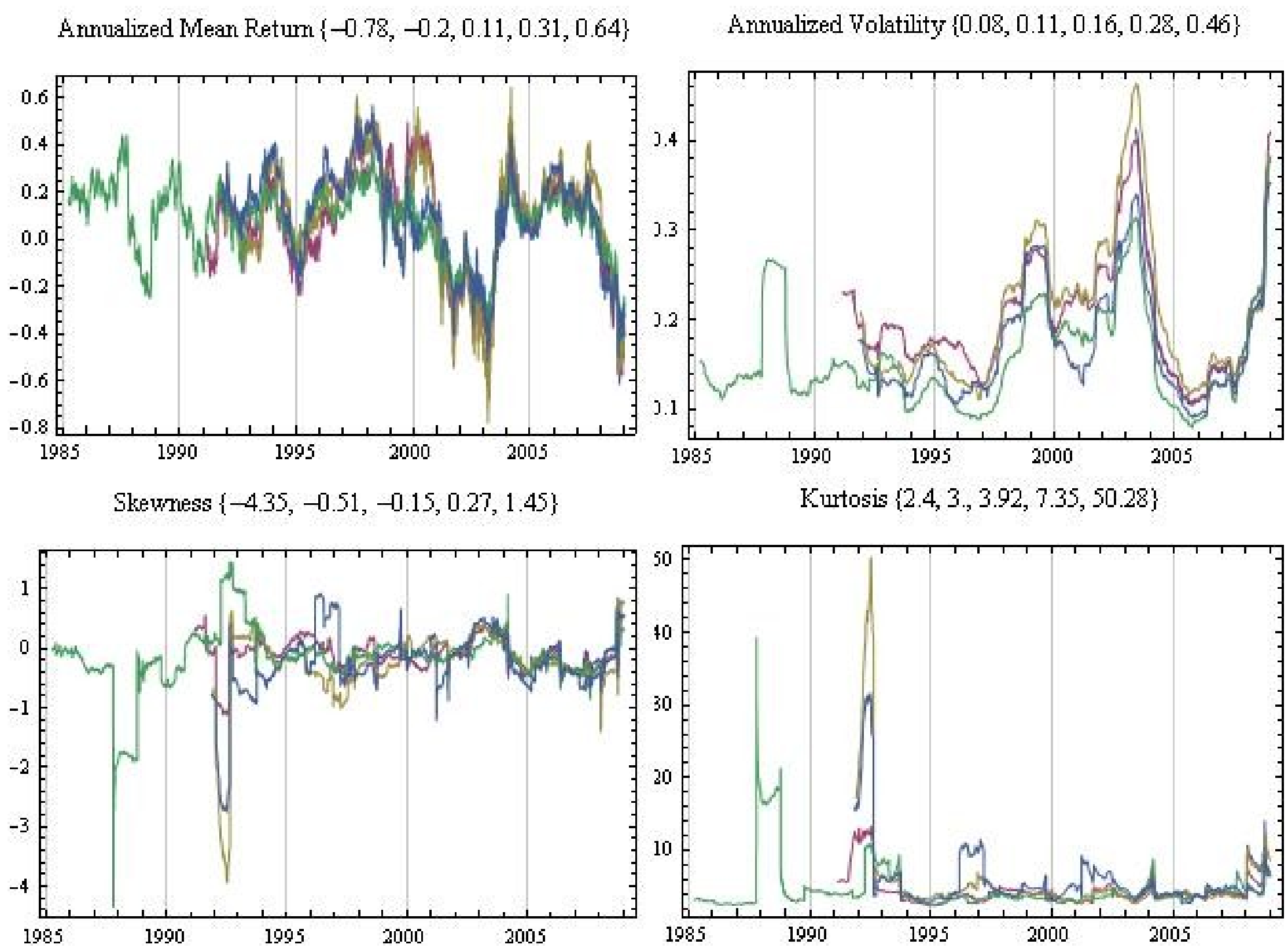

The ranges and dynamics of all four moments of the simulated model process are similar to those of the European market indices: both show rare but extreme jumps and changes in their moments.

The daily returns generated by the model are multiplied by a constant of 2.5 basis points so that the median of the volatility is comparable to that of the empirical values. One interpretation of this constant is that a single day of 2,048 up or down ticks can move no more than about 50 percent in either direction. In practice, neither the simulated model results nor the empirical values ever move that much.

Visually, even the patterns of returns look alike between the simulated and the empirical results, but this is a coincidence: different eras of European markets would show different patterns but the deterministic minimal model always looks the same.

**Figure 2**. **Model Moments.** The four graphs below show the rolling 252-day mean, volatility, skewness, and kurtosis for the model dynamics. The summary statistics in curly braces are the minimum, 10% quantile, median, 90% quantile, and maximum across all data points for all indices. Note that here too the moments differ considerably across time. The model is driven by a representative investor trading a single market asset based on its past price history. Specifically, these data points use a lookback window of 22 time periods. The process is cyclic and repeats after $2^{22}$ periods. To move from binary UP and DOWN changes to returns, the changes are first grouped into days of length 2,048, and scaled by 2.5 basis points so that the median volatility approximates the median empirical volatility of Figure 1 (though of course the skewness and kurtosis are unaffected by the rescaling).

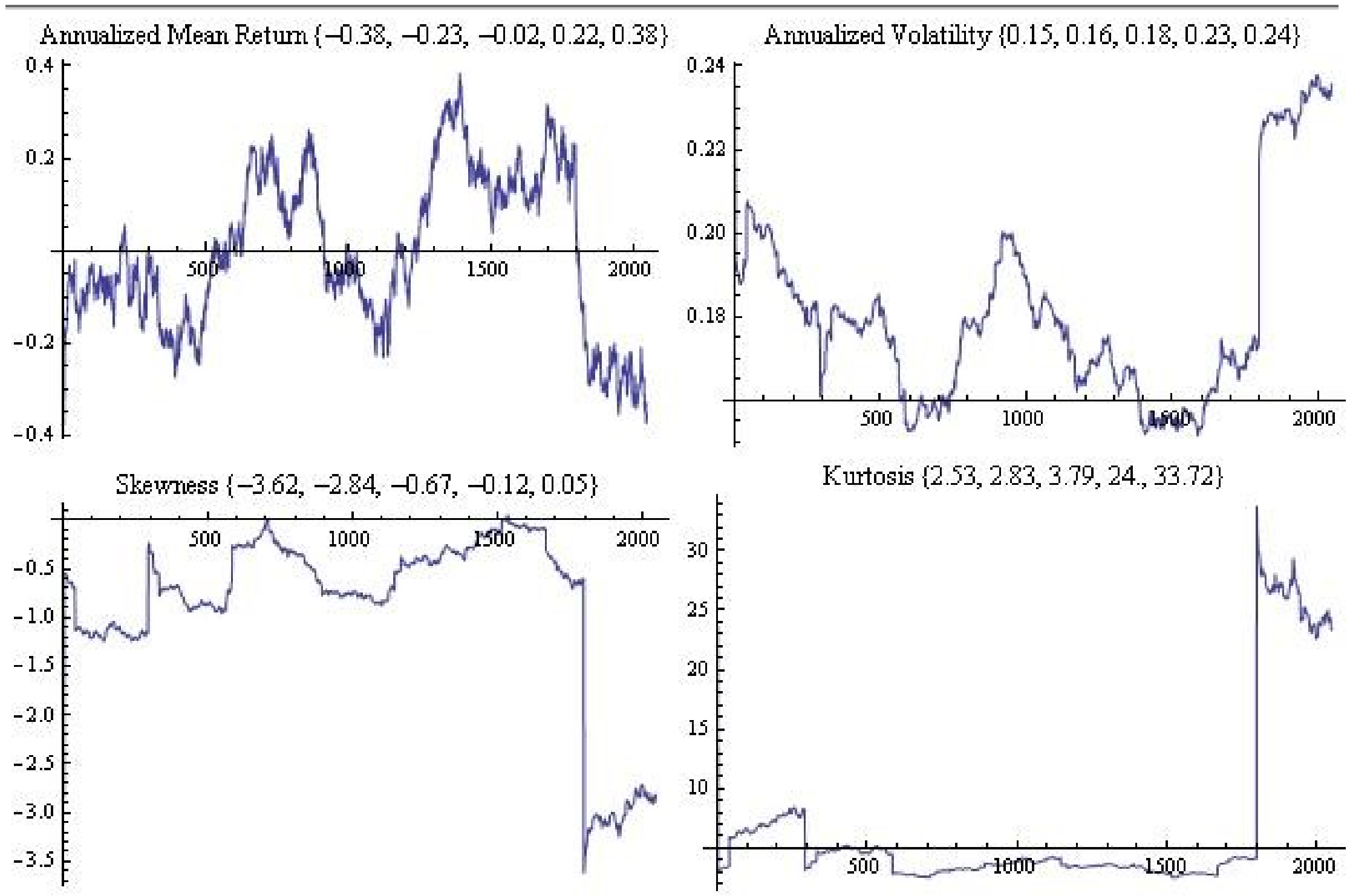

The key observation from these figures is that in both averages and dynamics, the model broadly appears to be a realistic depiction of the kind of path that a stock market could follow. We can therefore feel more comfortable in thinking that the results of regulation simulation on the model might be a realistic depiction of the kind of effect such regulations would have on the market if they were enacted.

Figures 3 and 4 show the effects of pricking bubbles, propping up crashes, or both, for trends of length $n$ = 6 and $n$ = 17, respectively, chosen to be representative of the change from short-term to long-term trend lengths.

**Figure 3**. **Simulated Returns with Regulation of Trends of Length 6**. These graphs show the rolling means, volatilities, skewnesses, and kurtoses of 256-day returns for the minimal model with a 22-tick lookback window and days comprised of 2,048 consecutive ticks under each possible regulatory environment. The statistics in curly braces are the minimum, 10% quantile, median, 90% quantile, and maximum across the associated regulatory environment.

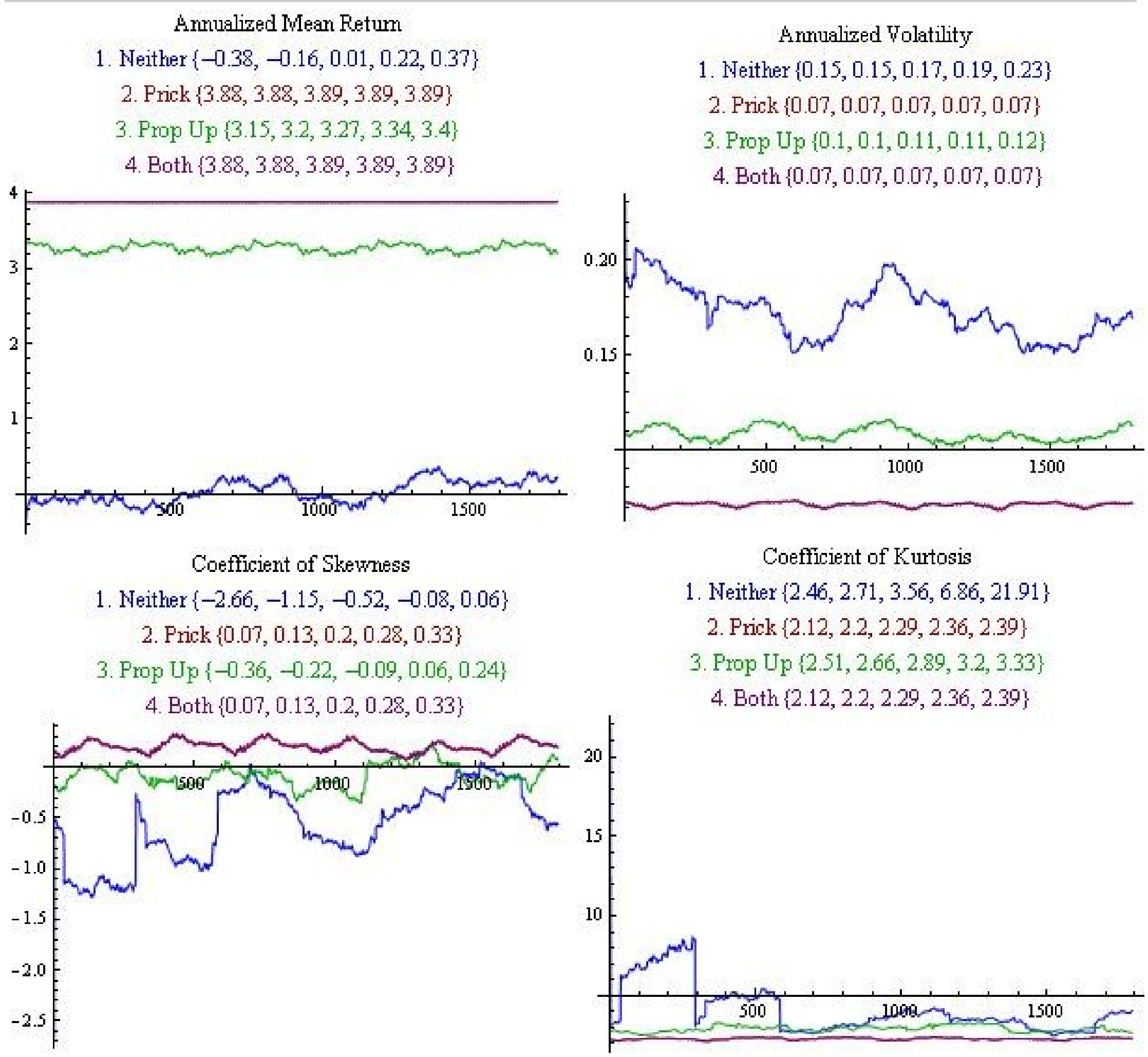

Each figure plots the rolling means, volatilities, skewnesses, and kurtoses of 256-day returns for the minimal model with a 22-tick lookback window and days comprised of 2,048 consecutive ticks under each possible regulatory environment.

Figure 3 shows that for short trends, any government intervention, whether propping up, pricking, or both, increases the mean return, decreases the volatility, and brings both the skewness and the kurtosis more in line with the Gaussian.

Figure 4 shows that for long trends, government intervention shifts the times when returns and volatilities are high or low without affecting the overall range, but it still eliminates the extremes of both skewness and kurtosis.

**Figure 4**. **Simulated Returns with Regulation of Trends of Length 17.** These graphs show the rolling means, volatilities, skewnesses, and kurtoses of 256-day returns for the minimal model with a 22-tick lookback window and days comprised of 2,048 consecutive ticks under each possible regulatory environment. The statistics in curly braces are the minimum, 10% quantile, median, 90% quantile, and maximum across the associated regulatory environment.

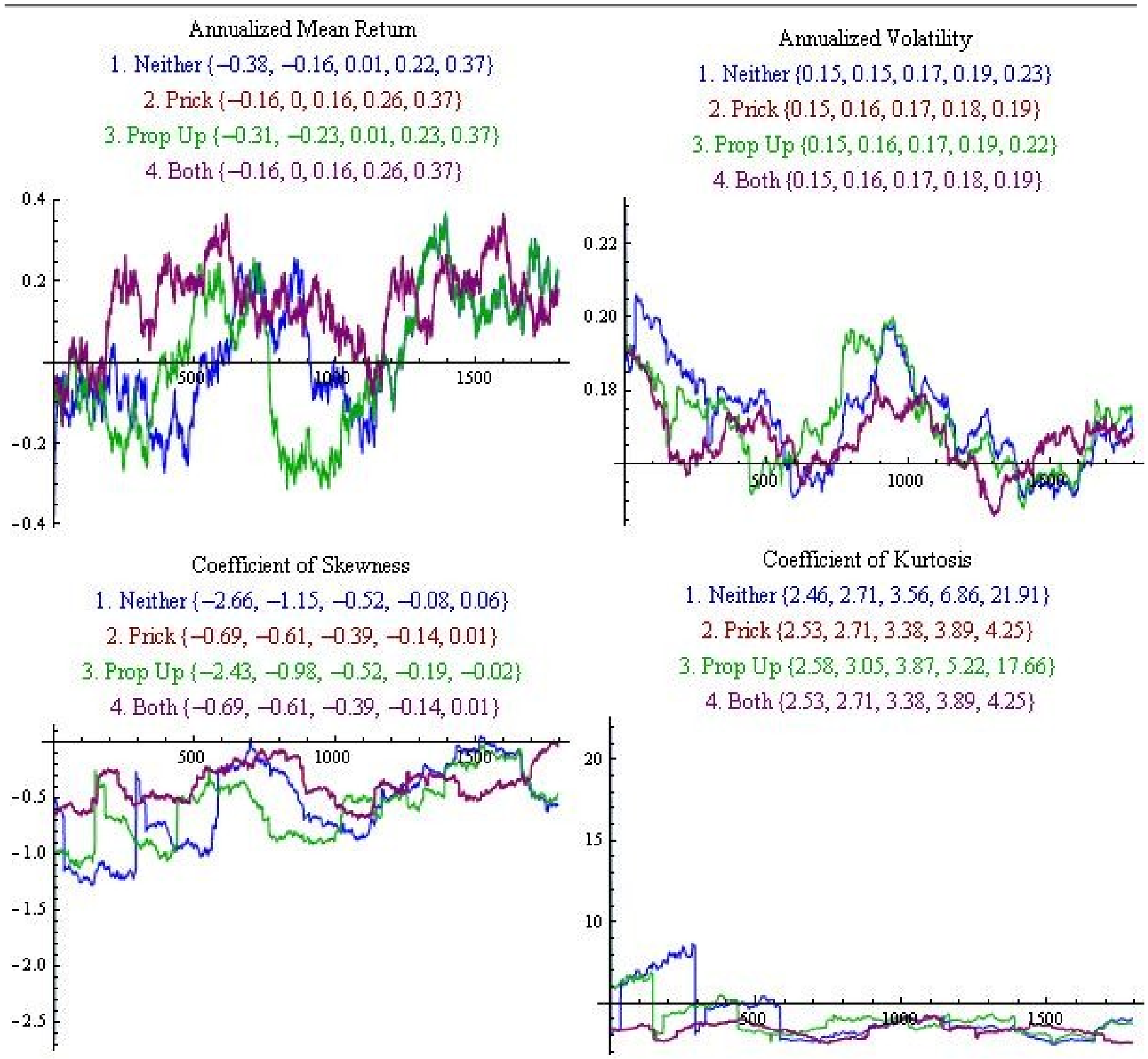

Table 1 lists the average annualized returns and volatilities and the maximum deviations from normality under the various regulatory regimes for trends ranging from $n$ = 2 through $n$ = 20. The maximum deviation from normality for skewness is the maximum absolute skewness observed during the simulations, while for kurtosis it is the maximum absolute value of the difference between the rolling sample annual kurtosis and three, the kurtosis of the normal distribution.

For example, for $n$ = 17, we can confirm from Figure 4 that the maximum deviation from normality for skewness (kurtosis) is 2.66, 0.69, 2.43, and 0.69 (18.91, 1.25, 14.66, and 1.25) under the regimes of no regulatory intervention, pricking only, propping up only, or both pricking and propping up, respectively.

Table 1 shows that the higher returns, lower volatilities, and more Gaussian distributions caused by regulatory intervention exist mainly for shorter trend lengths.

**Table 1. Regulated Price Processes.** This table lists the average of the rolling means and volatilities of 256-day returns for the minimal model with a 22-tick lookback window and days comprised of 2,048 consecutive ticks, as well as the maximum deviations from normality for the coefficients of skewness and kurtosis, for the assumption of no regulatory intervention at all (equivalent to a trend length of "0 or ∞", in blue), for regulatory pricking of bubbles only (first value in each column, in red), and for regulatory propping up of crashes only (second value in each column, in green).

| ***Trend Length*** | ***Average Annualized Mean Return*** | | ***Average Annualized Volatility*** | | ***Skewness Max Deviation from Normal*** | | ***Kurtosis Max Deviation from Normal*** | |
|---|---|---|---|---|---|---|---|---|
| 0 or ∞ | 1.43% | | 17.3% | | 2.66 | | 18.91 | |
| 2 | 3927% | 3055% | 0.4% | 4.4% | 0.64 | 0.81 | 1.61 | 1.10 |
| 3 | 2253% | 1708% | 1.5% | 4.8% | 0.04 | 0.45 | 1.32 | 0.61 |
| 4 | 1110% | 982% | 2.4% | 6.4% | 0.27 | 0.51 | 0.63 | 1.15 |
| 5 | 720% | 558% | 6.6% | 9.1% | 0.53 | 0.34 | 0.67 | 0.66 |
| 6 | 389% | 327% | 7.2% | 10.8% | 0.33 | 0.36 | 0.88 | 0.49 |
| 7 | 106% | 182% | 11.8% | 12.3% | 0.65 | 0.46 | 0.83 | 0.91 |
| 8 | 262% | 109% | 11.8% | 13.1% | 0.36 | 0.95 | 0.66 | 0.29 |
| 9 | 77% | 59% | 4.4% | 14.8% | 0.83 | 0.62 | 6.34 | 0.85 |
| 10 | 83% | 39% | 11.8% | 15.5% | 0.33 | 0.62 | 0.71 | 1.30 |
| 11 | -33% | 12% | 11.3% | 17.7% | 0.73 | 0.83 | 0.88 | 1.89 |
| 12 | 39% | 5% | 17.4% | 17.1% | 0.57 | 0.75 | 0.73 | 1.91 |
| 13 | 7% | 1% | 15.5% | 17.0% | 0.65 | 1.18 | 1.22 | 3.06 |
| 14 | -21% | 12% | 16.7% | 15.7% | 0.45 | 0.63 | 0.90 | 1.37 |
| 15 | 7% | 0% | 18.3% | 17.3% | 1.03 | 1.04 | 2.27 | 3.06 |
| 16 | 12% | 4% | 17.5% | 16.8% | 1.02 | 2.39 | 3.16 | 17.79 |
| 17 | 16% | 1% | 16.6% | 17.1% | 0.69 | 2.43 | 1.25 | 14.66 |
| 18 | 18% | 5% | 17.0% | 17.2% | 0.89 | 1.05 | 2.23 | 2.68 |
| 19 | 12% | 1% | 16.9% | 17.4% | 1.31 | 2.33 | 5.86 | 15.00 |
| 20 | 5% | 1% | 17.3% | 17.4% | 1.07 | 2.42 | 3.34 | 15.94 |

**A. Conditional on Bubble Pricking, Propping Up Crashes Doesn't Matter**

Observe from Figures 3 and 4 that the lines where the government only pricks bubbles (red) are always identical with those where the government both pricks bubbles and props up crashes (purple). This finding holds true for all trend lengths.

In other words, if the government pricks bubbles, it doesn't matter whether or not it also props up crashes; the resulting price process is exactly the same. Therefore, there are only three distinct regulatory possibilities: no intervention at all, only propping up crashes, and only pricking bubbles.

**B. The Results of Only Propping Up Crashes**

Let's compare the neutral (blue) lines with those where the government only props up (green). Until $n$ = 11, Table 1 shows that government prop ups result in higher mean returns, lower volatilities, less skew, and less kurtosis. In other words, government intervention by purchasing on consecutive dips makes the resulting returns appear more Gaussian. However, for longer trend lengths, as we can observe for example in Figure 4 for $n$ = 17, the regulated process can have lower mean, higher volatility, more extreme skew, and higher kurtosis than the non-regulated process. Also, as in Figures 3 and 4, the regulated process is typically less complex than the unregulated one: the regulated mean returns tend to form a much shorter cycle.

In short, propping up crashes makes the market look less like a real market.

**C. The Results of Only Pricking Bubbles**

Let's next compare the neutral (blue) lines with those where the government pricks bubbles (either purple or red). The conclusions are the same as between the neutral process and the prop-up process: government regulation with short trend lengths $n < 11$ results in a process that is more Gaussian, having higher mean returns, lower volatilities, less skew, and less kurtosis. As we can again see in Figure 4, at certain times, the regulated process can have lower mean, higher volatility, more extreme skew, and higher kurtosis than the non-regulated process. And again the regulated process is typically less complex than the unregulated one.

**D. Comparing the Results of Propping Up Crashes vs. Pricking Bubbles**

Finally, let's compare the process where the government only props up (green) to the process where the government only pricks bubbles (purple or red). Typically they are about the same but because of the lack of complexity by the bubble-pricing regulation, the comparisons are usually not very interesting. However, as a general rule, it appears the process resulting from bubble-pricking is even slightly closer to Gaussian than the process resulting from government propping up crashes.

## V. CONCLUSION

Should the government regulate the market? This paper seeks to address this broad question through simulation of a simple deterministic trading rule. The particular trading rule used as the basis of simulation is the unique minimal model that generates complexity in financial security prices. Furthermore, both the average moments of the generated process, as well as the time variation of those moments, are comparable to those empirically observed in European market indices. These observations give us comfort that simulations of regulation on the model may give us results whose intuitions also extend into the real world.

What do we discover? First, any regulation tends to simultaneously decrease the complexity of the market process and increase its similarity to a Gaussian process.

Second, propping up crashes and pricking bubbles are not symmetrical: if the government pricks bubbles, it does not matter whether or not it also props up crashes; the resulting price process is the same. However, if the government props up crashes, the price process could change again if the government decides to also prick bubbles.

Third, the degree to which regulations make the resulting price process closer to Gaussian depends on the trend length that the government aims to prick or prop up: the effect is more pronounced for shorter trends. In other words, to make the resulting price process seem more Gaussian, the government has to intervene more frequently.

Fourth, even when the regulations make the resulting price process more Gaussian, at any particular time point, the regulated process could be further from Gaussian than the original, non-regulated process. In other words, it matters when the regulation begins. An inopportune launch of new regulations could result in higher volatility, skew, and kurtosis, and lower mean returns, than would otherwise occur.

In sum, government intervention essentially precludes significant deviations by forbidding price movements that extend in a given direction for too long.

But what is the cost of such intervention? In the model considered here, it has implicitly been assumed to be zero in order to study the effects of the regulation itself, assuming it can be permanently funded as needed. Future research along this line could consider a government with limited resources that eventually must liquidate the positions that it establishes to offset bubbles and crashes. Other possible extensions include allowing for multiple traders, multiple assets, or multiple actions (e.g. buy, buy strongly, sell, sell strongly, or hold) within the same framework.